\documentclass{ws-ijmpe}

\begin{document}

\markboth{Jiangyong Jia}{$\pi^0$-h correlation in Cu+Cu}
\catchline{}{}{}{}{}
\title{$\pi^0$-h correlation in Cu+Cu at
$\sqrt{s_{NN}}$=200 GeV}

\author{\footnotesize Jiangyong Jia, for the PHENIX\footnote{For the full list of PHENIX authors and
acknowledgements, see Appendix 'Collaborations' of this volume}
Collaboration}

\address{Chemistry Department, State University of New York at Stony Brook,\\
Stony Brook, NY 11794-3400, USA}
\address{Physics Department, Brookhaven National Laboratory,\\
Upton, NY 11796, USA\\
jjia@bnl.gov} \maketitle

\maketitle

\begin{history}
\end{history}

\begin{abstract}
We present the results of the two-particle $\Delta\phi$
correlation between high $p_T$ $\pi^0$ and charged hadrons in
Cu+Cu collisions at $\sqrt{s_{NN}} = 200$ GeV. Clear away-side jet
signals are seen in central Cu+Cu collisions at all associated
hadron $p_T$, albeit suppressed by factor of 2 above 2 GeV/$c$.
This level is similar to the suppression seen for inclusive
hadrons at high $p_T$ ($>7$ GeV/$c$). However, this agreement is
argued to be the result of a cancellation between a flatter
away-side parton spectra due to the requirement of high $p_T$
triggers and a bigger energy loss due to a longer path length
traversed by the away-side jet.
\end{abstract}

\section{Introduction}
PHENIX have done a detailed mapping of the two-particle
correlations in centrality, $p_T$ and particle
types~\cite{Adler:2005ee,Adare:2006nr,Jia:2005ab,Grau:2005sm,Ajitanand:2006is,Adare:2006nn}.
Non-trivial modifications relative to p-p of the correlation
patterns are observed in the trigger $p_T$ ($p_T^A$) and partner
$p_T$ ($p_T^B$) below 4 GeV/$c$. There are strong evidences
suggesting that the $p_T$ dependence of the jet shape at the near-
and away-side is a consequence of the detailed interplay between
jet-quenching and response of the medium to the energy lost by
energetic jets~\cite{Jia:2007tu}. Additional complication (only
for jet yield) can be traced back to the soft contribution to the
triggers, which are used to normalize jet pair rate into
per-trigger yield ($Y_{\rm{jet}}$)~\cite{Jia:2007tu}. For example,
imagine we try to correlate hadrons in 2-3 GeV/$c$ ($p_T^1$) with
those 5-10 GeV/$c$ ($p_T^2$). Since
$R_{\rm{AA}}(p_T^1)>R_{\rm{AA}}(p_T^2)$, arguably due to
recombination contribution, one expects $I_{\rm{AA}}$ (ratio of
$Y_{\rm{jet}}$ in A+A to that in p+p) for using $p_T^1$ as trigger
and $p_T^2$ as partner is lower than the other way
around~\footnote{i.e.
$R_{\rm{AA}}(p_T^1)I_{\rm{AA}}(p_T^1,p_T^2)=R_{\rm{AA}}(p_T^2)I_{\rm{AA}}(p_T^2,p_T^1)$~\cite{Jia:2006im}
. Note $I_{\rm{AA}}$ is function of trigger and partner $p_T$:
$I_{\rm{AA}}(p_{T,\rm{trigger}},p_{T,\rm{partner}})$.}.

In this analysis, we present results using high $p_T$ $\pi^0$ as
triggers. These triggers comes dominantly from jet fragmentation,
thus provide a cleaner physical interpretation of the per-trigger
yield. In terms of $N_{\rm{part}}$, Cu+Cu collisions cover from
peripheral to 30-40\% central Au+Au collisions. $N_{\rm{part}}$
can be determined with better precision that in Au+Au, thus allows
a more detailed mapping of the centrality dependence of the onset
of the jet quenching. Additionally, systematic error from on the
jet yield due to elliptic flow is much more reduced in Cu+Cu due
to a smaller combinatoric background level.

\section{Analysis and results}
The measurement was carried with 5-10 GeV/$c$ $\pi^0$ as triggers
and charged hadrons in several ranges in 0.4-10 GeV/$c$ as
partners. The per-trigger jet yield is obtained
via~\cite{Adler:2005ee,Jia:2004sw,Adler:2005ad},
\begin{eqnarray}
\label{eq:core1}\nonumber Y_{\rm{jet}} = \frac{\int d\Delta\phi
N^{\rm{mix}}}{2\pi
N_{t}\varepsilon_B}\left(\frac{N^{\rm{same}}(\Delta\phi)}{N^{\rm{mix}}(\Delta\phi)}-
\xi\left(1+2v_2^{t} v_2^{a} \cos2\Delta \phi\right)\right)
\end{eqnarray}
where $N_{A}$ is the number of triggers, $\varepsilon_B$ is the
single particle efficiency for partners in the full azimuth and
$\left|\eta\right| < 0.35$; $N^{\rm{same}}(\Delta\phi)$ and
$N^{\rm{m}}(\Delta\phi)$ are pair distributions from the same- and
mixed-events, respectively. Mixed-event pairs are obtained by
selecting partners from different events with similar centrality
and vertex. The $\varepsilon_B$ values include detector acceptance
and reconstruction efficiency, with an uncertainty of $\sim 10$\%,
evaluated similar to ~\cite{Adler:2003au}. $\xi$ is a
normalization factor which is the ratio of the combinatorics pairs
in the same event to those in the mixed event. $\xi$ is typically
bigger but very close to 1.

\begin{figure}[h]
\epsfig{file=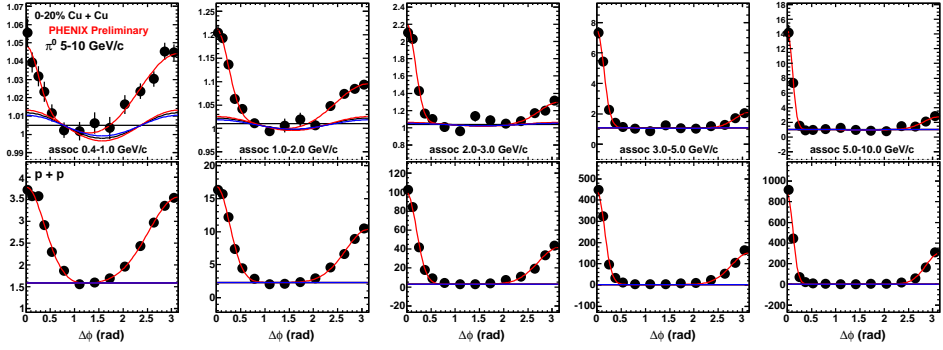,width=1.0\columnwidth}
\caption{\label{fig:cucucf} The correlation function in Cu+Cu
(upper panels) and p+p (bottom panels) for 5-10 GeV/$c$ $\pi^0$
and various partner $p_T$ ranges within 0.4-10 GeV/$c$.}
\end{figure}

\begin{figure}[h]
\epsfig{file=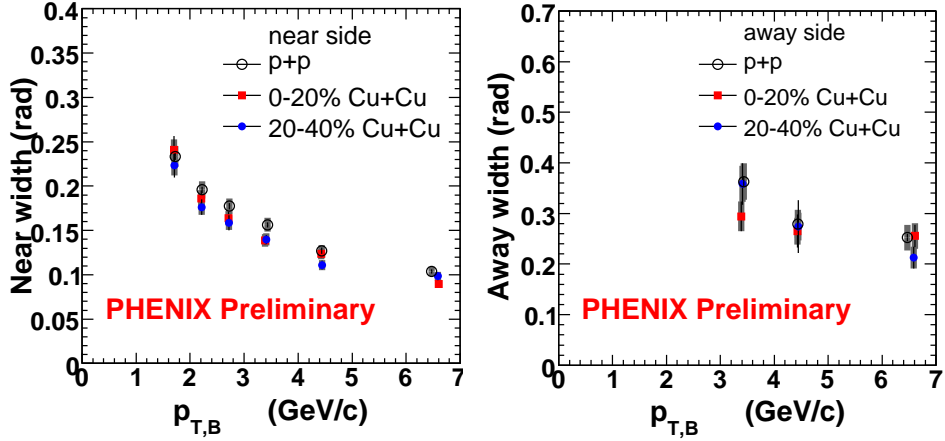,width=1.0\columnwidth}
\caption{\label{fig:cucuwidth} The near-side and away-side jet
gauss width vs. $p_T^B$ for p+p, 0-20\% and 20-40\% Cu+Cu
collisions.}
\end{figure}

Fig.\ref{fig:cucucf} show the correlation function, $CF=
N^{\rm{same}}(\Delta\phi)/N^{\rm{mix}}(\Delta\phi)$, in 0-20\%
Cu+Cu compared with that in p+p. A clear away-side excess can be
seen in all partner $p_T$ ranges. No concave shape is expected
even after the $v_2$ background subtraction. At partner $p_T>2$
GeV/$c$, the away-side can be well described by a gauss function.
Fig.\ref{fig:cucuwidth} shows the near-side and away-side jet
width as function of partner $p_T$ for p+p, 0-20\% and 20-40\%
Cu+Cu collisions. No significant differences can be seen between
Cu+Cu and p+p within systematic errors, suggesting partners come
mostly from jet fragmentation in the $p_T$ range under
consideration ($0.4 <p_T^B<10$ GeV/$c$  at the near side and
$p_T^B>$2 GeV/$c$ at the away side).

\begin{figure}[ht]
\epsfig{file=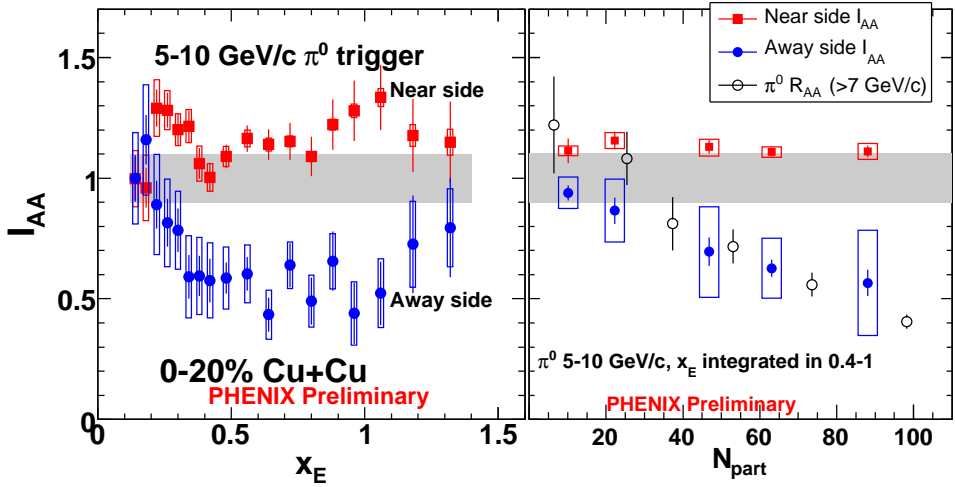,width=1\columnwidth}
\caption{\label{fig:xe} a) The $I_{\rm{AA}}$ as function of $x_E$
in p+p and Au+Au for near and away-side; b) The $I_{\rm{AA}}$ for
yield integrated in 0.4-1 as function of $N_{\rm{part}}$, compared
with suppression factor for high $p_T$ $\pi^0$.}
\end{figure}

Comparing the top and the bottom panels of Fig.\ref{fig:cucucf},
one notices that there is a larger asymmetry between the near- and
away-side in Cu+Cu than that in p+p. To quantify the medium
modifications, we extract the $Y_{\rm{jet}}$ for Cu+Cu and p+p and
construct the $I_{\rm{AA}}$ as function of $x_E$ ($x_E =
p_{T,A}/p_{T,T}\cos(\Delta\phi)$). The results for 0-20\% central
Cu+Cu collisions are shown in left panel of Fig.\ref{fig:xe}.
$I_{\rm{AA}}$ at the near side is consistent with 1 in the full
$x_E$ range of 0.1-1.4; the away-side $I_{\rm{AA}}$ starts at
slightly above 1 at small $x_E$, and gradually decreases towards
larger $x_E$, at $x_E>0.4$ the suppression value approaches a
constant of 0.5 up to $x_E\sim1$. This constant behavior is also
seen by STAR in Au+Au collisions~\cite{Adams:2006yt}, where the
level of suppression was found to be similar to single particle
$R_{\rm{AA}}$.

The right panel of Fig.\ref{fig:xe} shows the integrated
$I_{\rm{AA}}$ in $0.4<x_E<1$ as function $N_{\rm{part}}$,
comparing with the integrated $R_{\rm{AA}}$ for the single
particle. Assuming $\langle z\rangle\sim0.7$ for leading $\pi^0$s,
the $p_T$ of the original jets should be around 5/0.7=7 GeV/$c$.
Thus the $I_{\rm{AA}}$ of the away-side should be directly
comparable to the single particle at $p_T>7$ GeV/$c$. In reality,
since the $\pi^0$ $R_{\rm{AA}}$ is flat at $p_T>4$ GeV/$c$, the
exact values of the energy of the jets for high $p_T$ $\pi^0$s are
not important.

Fig.\ref{fig:xe}b indicates that the near-side $I_{\rm{AA}}$ is
around one in all centralities, consistent with surface emission
picture. In contrast, away-side $I_{\rm{AA}}$ shows a suppression
that has a similar centrality dependence in $N_{\rm{part}}$
relative to single particle suppression. This is rather surprising
given that the away-side jet travels more medium than the single
particles in the naive jet absorption picture~\cite{Drees:2003zh}.
Suggesting that the simple geometrical bias argument is too naive.
In jet quenching picture, the suppression is due to the energy
degradation of the high $p_T$ jets. The suppression factor
$R_{\rm{AA}}$ depends on both the energy loss as well as on the
input parton spectra shape. For the single particle, the expected
$N_{\rm{binary}}$ scaled p+p spectra have a typical power-law
shape with a power of 8. In the di-hadron correlation, the
away-side spectra associated with the leading particles have much
flatter distribution with a much smaller power. Thus for the same
amount of energy loss for single jet and away-side jet, the
suppression level observed for $I_{\rm{AA}}$ could be smaller than
the that for $R_{\rm{AA}}$.

\begin{figure}[ht]
\epsfig{file=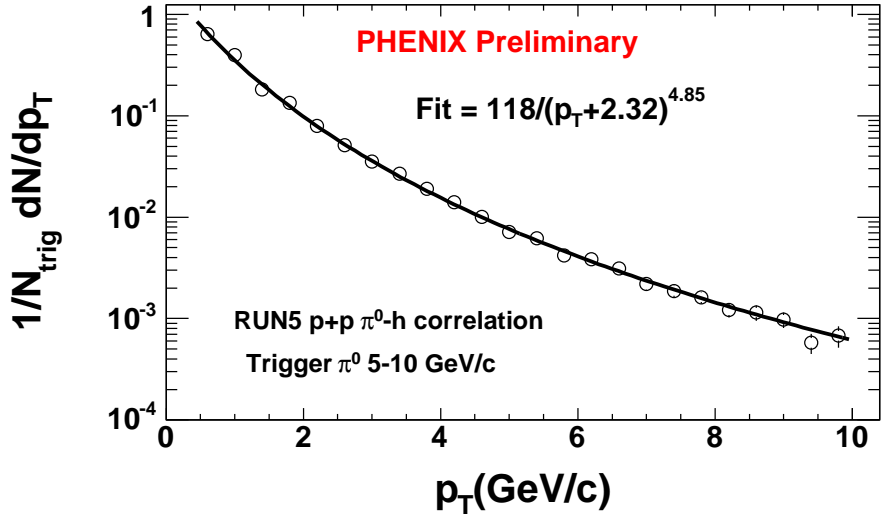,width=1.0\columnwidth}
\caption{\label{fig:pp} The RUN5 p+p away-side jet yield
associated with 5-10 GeV/c trigger $\pi^0$ and the corresponding
modified power-law fit}
\end{figure}

To illustrate this idea, we follow the prescription in
Ref.~\cite{Adler:2006bw}. The fractional energy loss
$S_{\rm{loss}}$ is related to $R_{\rm{AA}}$ as:
\begin{eqnarray}
S_{\rm{loss}} = 1 - R_{\rm{AA}}(p_T)^{1/(n-1)} \quad or \quad
R_{\rm{AA}} = (1-S_{\rm{loss}})^{n-1}
\end{eqnarray}
The power ``n'' is $7.1$ for single particle spectra in $dN/dp_T$.
The power for the away-side associated spectra can be determined
from RUN5 $p+p$ data in Fig.\ref{fig:pp} (see also RUN3 p+p data
results~\cite{Adler:2005ad}) via a modified power-law fit
:$dN/dp_T=A/(p_T+p_0)^{n}$. The power n is found to be about 4.8.

If we assume the away-side jet have same fraction energy loss as
single particles, we have:
\begin{eqnarray}
S_{\rm{loss}} = 1 - R_{\rm{AA}}(p_T)^{1/(n_{R}-1)} =1 -
I_{\rm{AA}}(p_T)^{1/(n_{I}-1)}.
\end{eqnarray}
$R_{\rm{AA}} =0.2$ in central Au+Au collisions would lead to
$I_{\rm{AA}} = R_{\rm{AA}}^{(n_I-1)/(n_{R}-1)}= 0.37$, indicating
a much smaller suppression than that for $R_{\rm{AA}}$. On the
other hand, if we require $I_{\rm{AA}}= R_{\rm{AA}}$, then the
away hadron energy loss fraction would be $S^{I}_{\rm{loss}} = 1 -
I_{\rm{AA}}(p_T)^{1/(n_{I}-1)} = 0.345$, much bigger than the
single hadron energy loss fraction $S^{R}_{\rm{loss}}=0.23$, as
expected (about 50\% more energy loss). If we apply this trick to
central Cu+Cu collisions, and assumes $I_{\rm{AA}} = R_{\rm{AA}}
=0.5$, the fractional energy loss for away-side jet would be
$S^{R}_{\rm{loss}} = 0.167$. This value is significantly larger
than the value for inclusive hadrons of $S^{I}_{\rm{loss}} =
0.167$.

In summary, high $p_T$ $\pi^0-h$ correlations are studied in Cu+Cu
and p+p collisions. Clear, relatively unmodified jet peaks are
seen in rather low partner $p_T$ in central Cu+Cu collisions,
albeit suppressed by factor of 2 that is similar to that for
inclusive hadrons. These observations suggest that the high $p_T$
correlation is dominated by fragmentation of the jet that survives
the medium. The similar suppression level observed for
$I_{\rm{AA}}$ relative to $R_{\rm{AA}}$ is the combined results of
a larger energy loss and a flatter parton spectra.


\begin{thebibliography}{0}

\bibitem{Adler:2005ee}
  S.~S.~Adler {\it et al.}
  Phys.\ Rev.\ Lett.\  {\bf 97}, 052301 (2006)
\bibitem{Adare:2006nr}
  A.~Adare  {\it et al.}
  nucl-ex/0611019.
\bibitem{Jia:2005ab}
  J.~Jia  [PHENIX Collaboration],
  AIP Conf.\ Proc.\  {\bf 828}, 219 (2006)
\bibitem{Grau:2005sm}
  N.~Grau  [PHENIX Collaboration],
  Nucl.\ Phys.\  A {\bf 774}, 565 (2006)
\bibitem{Ajitanand:2006is}
  N.~N.~Ajitanand  [PHENIX Collaboration],
  Nucl.\ Phys.\  A {\bf 783}, 519 (2007); C.~Zhang, QM2006
  proceedings.
\bibitem{Adare:2006nn}
  A.~Adare  [PHENIX Collaboration],
  arXiv:nucl-ex/0611016.
\bibitem{Jia:2007tu}
  J.~Jia,
  arXiv:nucl-ex/0702048.

\bibitem{Jia:2006im}
  J.~Jia,
  AIP Conf.\ Proc.\  {\bf 842}, 116 (2006)
  [arXiv:nucl-ex/0601023].
\bibitem{Jia:2004sw}
  J.~Jia,
  J.\ Phys.\ G {\bf 31}, S521 (2005)
  [arXiv:nucl-ex/0409024].

\bibitem{Adler:2005ad}
  S.~S.~Adler {\it et al.}
  Phys.\ Rev.\ C {\bf 73}, 054903 (2006)
\bibitem{Adler:2003au}
  S.~S.~Adler {\it et al.}
  Phys.\ Rev.\ C {\bf 69}, 034910 (2004)
\bibitem{Adams:2006yt}
  J.~Adams {\it et al.}  [STAR Collaboration],
  arXiv:nucl-ex/0604018.

\bibitem{Adler:2006bw}
  S.~S.~Adler  [PHENIX Collaboration],
  arXiv:nucl-ex/0611007.


%
%
\bibitem{Drees:2003zh}
  A.~Drees, H.~Feng and J.~Jia,
  Phys.\ Rev.\  C {\bf 71}, 034909 (2005)
\end{thebibliography}
\end{document}